\documentclass[aoas,preprint]{imsart}

\RequirePackage[OT1]{fontenc}
\RequirePackage{amsthm,amsmath}
\RequirePackage{natbib}

\usepackage{bm}
\usepackage{amssymb}
\usepackage{graphics, graphicx}
\usepackage{float}
\usepackage[usenames,pdftex,dvipsnames]{color}
\usepackage{bbm}
\usepackage{mathtools}
\usepackage{subfig}

\usepackage[usenames,dvipsnames]{color}
\usepackage{color,soul}
\usepackage{xcolor,framed}
\definecolor{shadecolor}{HTML}{FFF200}
\usepackage{ifthen}
\newboolean{isclean}
\setboolean{isclean}{false}
\ifthenelse{\boolean{isclean}}{

}{}

\DeclarePairedDelimiter\ceil{\lceil}{\rceil}

\startlocaldefs
\numberwithin{equation}{section}
\theoremstyle{plain}

\endlocaldefs

\begin{document}

\begin{frontmatter}
\title{Bayesian Landmark-based Shape Analysis of Tumor Pathology Images}
\runtitle{Bayesian landmark-based shape analysis}
\thankstext{T1}{ Address correspondence to qiwei.li@utdallas.edu}

\begin{aug}
\author{\fnms{Cong} \snm{Zhang}\thanksref{m1}\ead[label=e1]{cong.zhang3@utdallas.edu}},	
\author{\fnms{Guanghua} \snm{Xiao}\thanksref{m2}\ead[label=e2]{guanghua.xiao@utsouthwestern.edu}},
\author{\fnms{Chul} \snm{Moon}\thanksref{m3}\ead[label=e5]{chulm@mail.smu.edu}},
\author{\fnms{Min} \snm{Chen}\thanksref{m1}\ead[label=e3]{mchen@utdallas.edu}},
\and
\author{\fnms{Qiwei} \snm{Li}\thanksref{m1,T1}\ead[label=e4]{qiwei.li@utdallas.edu}
}

\runauthor{Zhang et al.}

\affiliation{The University of Texas at Dallas\thanksmark{m1}, The University of Texas Southwestern Medical Center\thanksmark{m2}, and Southern Methodist University\thanksmark{m3}}

\address{Cong Zhang, Min Chen, and Qiwei Li\\
	800 W Campbell Rd\\
	Mathematical Sciences, FO 35\\
	The University of Texas at Dallas\\
	Richardson, TX 75080, United States\\
	\printead{e1}\\
\printead{e3}\\
\printead{e4}}

\address{Chul Moon\\
	3225 Daniel Ave\\
	104 Heroy Science Hall\\
	Southern Methodist University\\
	Dallas, TX 75275, United States\\
	\printead{e5}\\}

\address{Guanghua Xiao\\
5323 Harry Hines Blvd\\
Quantitative Biology Research Center, Suite H9.124\\
The University of Texas Southwestern Medical Center\\
Dallas, TX 75390, United States\\
\printead{e2}}

\end{aug}

\begin{abstract}
Medical imaging is a form of technology that has revolutionized the medical field in the past century. In addition to radiology imaging of tumor tissues, digital pathology imaging, which captures histological details in high spatial resolution, is fast becoming a routine clinical procedure for cancer diagnosis support and treatment planning. Recent developments in deep-learning methods facilitate the segmentation of tumor regions at almost the cellular level from digital pathology images. The traditional shape features that were developed for characterizing tumor boundary roughness in radiology are not applicable. Reliable statistical approaches to modeling tumor shape in pathology images are in urgent need. In this paper, we consider the problem of modeling a tumor boundary with a closed polygonal chain. A Bayesian landmark-based shape analysis (BayesLASA) model is proposed to partition the polygonal chain into mutually exclusive segments to quantify the boundary roughness piecewise. Our fully Bayesian inference framework provides uncertainty estimates of both the number and locations of landmarks. The BayesLASA outperforms a recently developed landmark detection model for planar elastic curves in terms of accuracy and efficiency. We demonstrate how this model-based analysis can lead to sharper inferences than ordinary approaches through a case study on the $246$ pathology images from $143$ non-small cell lung cancer patients. The case study shows that the heterogeneity of tumor boundary roughness predicts patient prognosis ($p$-value $<0.001$). This statistical methodology not only presents a new model for characterizing a digitized object's shape features by using its landmarks, but also provides a new perspective for understanding the role of tumor surface in cancer progression.

\end{abstract}



\end{frontmatter}

\section{Introduction}\label{introduction}
\begin{figure}[h]
	\begin{center}
		\includegraphics[width=1.0\linewidth]{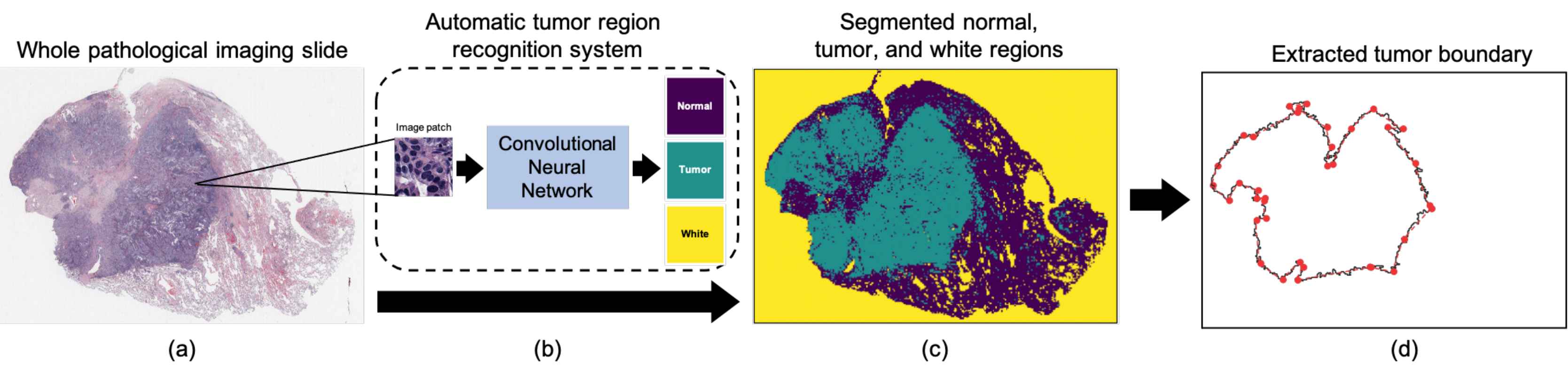}
	\end{center}
	\caption{Illustration of the pipeline: (a) The whole pathological imaging slide from a lung cancer patient (the median size of those slides in the analyzed NLST dataset is $24,244\times19,261$ pixel); (b) The convolutional neural network that predicts that type of each $300\times300$ pixel image patch into three categories: normal (purple), tumor (green), and white (yellow) \citep{wang2018comprehensive}; (c) The resulting ternary image corresponding to the original grayscale pathology image as shown in (a); (d) The tumor boundary extracted from the tumor (green) region as shown in (c) and the identified landmarks (in red) by the proposed BayesLASA.}
	\label{Figure 1}
\end{figure}

Cancer, a group of diseases characterized by uncontrolled tumor cell growth, is a major leading cause of death worldwide \citep{wang2016global}. Imaging studies using computed tomography (CT) colonography, magnetic resonance imaging (MRI), and positron emission tomography (PET)/CT colonography is proven to be valuable in the evaluation of patients for the screening, staging, surveillance, and treatment planning of cancer \citep{kijima2014preoperative,mohammadzadeh2015advances}. In addition to using these radiographic imaging techniques, hematoxylin and eosin (H\&E)-stained pathology imaging (see an example in Figure \ref{Figure 1}(a)) is fast becoming a routine procedure in clinical diagnosis and prognosis of various malignancies \citep{niazi2019digital}.

Feature extraction is an essential part of the radiomics workflow, which serves as the bridge between medical imaging to clinical endpoints \citep{gillies2016radiomics,lambin2017radiomics}. The two most widely used types of features are texture and shape \citep{bianconi2018evaluation}. Texture features ranges from summary statistics such as the mean, standard deviation, skewness, and kurtosis of the gray-level distribution to the second or higher-order statistics-based co-occurrence and run-length matrices \citep{larroza2016texture}. 
In complement to textural features, shape features are often extracted in radiomic analysis to describe tumor aggressiveness. This is because spiculated margins (or ``ill-defined borders") indicate the invasion of tumor cells into surrounding tissues \citep{edge2010american}. In contrast, benign tumors usually have well-defined margins \citep{razek2011soft}. In general, there are three kinds of shape features: 1) geometrical descriptors based on spatial moments \cite[see e.g.][]{shen1994application,pohlman1996quantitative,rangayyan1997measures}; 2) topological descriptors based on Euler characteristics \cite[see e.g.][]{crawford2020predicting}; 3) boundary descriptors based on radial distance measures \cite[see e.g.][]{kilday1993classifying,bruce1999effects,georgiou2007multi, li2013enhanced, rahmani2016assessment,sanghani2019evaluation} and fractal dimensions \cite[see e.g.][]{bru2008fractal,klonowski2010simple,rajendran2019initial}. However, extracting clinically meaningful imaging features remains a challenging problem in diagnostic medicine, particularly from digital pathology images that differ vastly from radiology images with regard to their spatial resolution \citep{sadimin2012pathology}.

Current studies of pathology image analysis mainly focus on morphological texture features. For instance, \cite{Tabesh2007} aggregated color, texture, and morphometric cues at the global and histological object levels predict prostate cancer's malignancy level. Both \cite{Yu2016} and \cite{Luo2016} used a large number of objective descriptors (e.g. cell size, shape, distribution of pixel intensity in the cells and nuclei, texture of the cells and nuclei, etc.) extracted by CellProfiler \citep{Carpenter2006,Kamentsky2011} to predict lung cancer prognosis. \cite{Yuan2012} integrated morphological texture features with histopathology and genomics information to predict breast cancer patient survival outcomes. However, these imaging data, which capture tumor histomorphological details in high resolution, still leave unexplored more undiscovered knowledge. Notably there is a lack of rigorous statistical methods to derive tumor shape features due to the high complexity of pathology imaging data.

Although the use of radial distance-based shape features, such as the tumor boundary roughness and zero-crossing count, has shown success in classifying malignant and benign breast tumors \citep{kilday1993classifying,georgiou2007multi,li2013enhanced,rahmani2016assessment} and predicting brain tumor prognosis \citep{sanghani2019evaluation,vadmal2020mri} from radiology (e.g. CT and MRI) images, we found that they performed poorly to characterize tumor border irregularity in pathology images. This might be due to two issues: 1) those features were developed for radiology images with a low spatial resolution (i.e. at anatomy level), which might not be suitable for high-resolution pathology images (i.e. at cell level); 2) those features assumed homogeneous irregularity across the tumor boundary, while different tumor boundary segments can show distinct morphological profiles. 
To address the first issue, \cite{wang2018comprehensive} have developed a deep convolutional neural network to classify image patches in a pathology image into three categories: normal, tumor, and white (see Figure \ref{Figure 1}(b)). This system, which transforms a grayscale pathology image into a ternary image (see Figure \ref{Figure 1}(c)), enables us to recognize tumor regions at almost the cellular level from digital pathology images at large scale. To overcome the challenge resulted from the second issue, new statistical methods that account for heterogeneous boundary roughness are required to analyze tumor shapes. 

Our basic idea is to use summary statistics of the piecewise roughness measurements to characterize the heterogeneity, where segments are partitioned by a set of \textit{landmarks} that approximately reconstruct the tumor shape (see Figure \ref{Figure 1}(d)). In practice, radiologists usually annotate landmarks for shape analysis \citep{houck2020comparison}. However, this process is laborious, tedious, and subject to errors. Meanwhile, some automatic landmark identification approaches were developed based on global convexity \citep{subburaj20083d,zulqarnain2015shape} or local curvature \citep{liu2012landmark}. However, those methods have been challenged by low robustness and infeasible uncertainty assessment due to a lack of underlying statistical models. The identification of landmarks of a shape has been a primary focus in shape analysis. \cite{domijan2005bayesian} presented a model-based approach without considering shape-preserving transformations. Recently, \cite{strait2019automatic} proposed a Bayesian model to detect the number and locations of landmarks using square-root velocity function representation under the elastic curve paradigm. However, the high computational cost may hinder its application in analyzing medical images, particularly high-resolution pathology images.

In this paper, motivated by the emerging needs of redefining shape features for tumor pathology images, we develop a Bayesian landmark detection model that can be served as a novel model-based approach to characterize the heterogeneity of tumor boundary roughness. Considering the sequence of boundary pixels of a tumor region as a closed polygonal chain, we aim to identify a set of landmarks in a simple polygon (i.e. tumor region). Those landmarks partition the whole polygonal chain (i.e. tumor boundary) into mutually exclusive segments. Our fully Bayesian inference framework provides uncertainty estimates of both the number and locations of landmarks. Compared with two alternative approaches, the proposed Bayesian LAndmark-based Shape Analysis (BayesLASA) performs well in simulation studies in terms of landmark identification accuracy and computational efficiency. We also conduct a case study on a large cohort of lung cancer pathology images. The result shows that the skewness and kurtosis of piecewise-defined roughness measurements, based on either the conventional surface profiling or hidden Markov modeling approach, are associated with patient prognosis ($p$-value $<0.001$). In this study, the proposed methodology not only delivers a new perspective for understanding the role of tumor shape/boundary in cancer progression, but also provides a refined statistical tool to characterize an object's shape features, while preserving the heterogeneity of its boundary roughness.

The remainder of the paper is organized as follows. Section \ref{model} introduces the proposed BayesLASA model and discusses the parameter structure as well as the prior formulation. Section \ref{model fitting} briefly describes the Markov chain Monte Carlo (MCMC) algorithm and the resulting posterior inference for the parameter of interest, the landmark indicators. In Section \ref{simulation}, we evaluate the BayesLASA on simulated data, comparing with two alternative approaches. Section \ref{application} consists of a lung cancer case study with several downstream analyses. Section \ref{conclusion} concludes the paper with remarks on future directions.

\section{Model}\label{model}
In this section, we introduce a parametric model for detecting landmarks in a polygonal chain. Although the model is applicable to any two-dimensional closed or open polygonal chains, it can be easily extended for polygonal chains in any dimension. 

\subsection{Observed Data: A Polygonal Chain}
\begin{figure}[h]
	\begin{center}
		\includegraphics[width=1.0\linewidth]{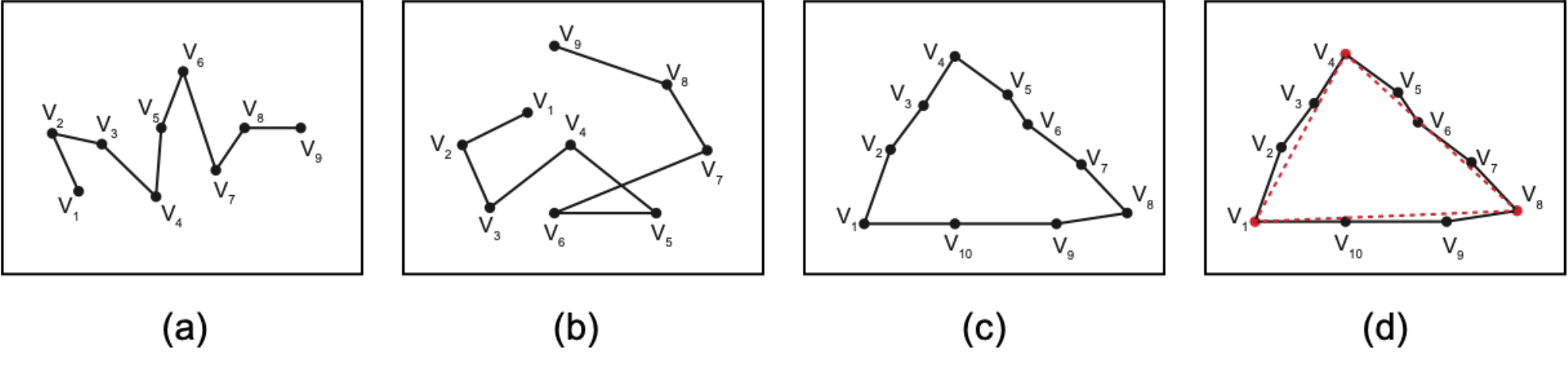}
	\end{center}
	\caption{Examples of different chains defined in this paper: (a) a simple open polygonal chain with $n=9$ vertices; (b) a self-intersecting polygonal chain with with $n=9$ vertices; (c) a simple closed polygonal chain with $n=10$ vertices; (d) the closed polygonal chain as shown in (c) and an associated landmark chain (the red dashed lines) with $K=3$ landmarks (the red vertices). The parameterization of landmarks can be $\bm{\gamma}=(1,0,0,1,0,0,0,1,0,0)$ or $\bm{z}=(1,1,1,2,2,2,2,3,3,3)$.}
	\label{Figure 2}
\end{figure}

The tumor boundary in a pathology image can be abstracted into a sequence of almost equally spaced discretization points (i.e. image pixels or patches), which could be considered as a \textit{closed polygonal chain}. In geometry, a polygonal chain is a connected series of line segments, each of which is a part of a line that is bounded by two distinct endpoints. Mathematically, a polygonal chain ${P}$ is a discretized curve specified by a sequence of vertices $(V_1,\ldots,V_n)$. Each vertex $V_i,i=1,\ldots,n$ can be represented by its coordinate $(x_i,y_i)\in\mathbb{R}^2$ in a two-dimensional Cartesian plane. Note that although we focus on planar polygonal chains here, the proposed method can be easily extended to a general case of $\mathbb{R}^k,k\ge3$. A \textit{simple polygonal chain} is one in which only consecutive segments intersect at their endpoints, while its opposite is a self-intersecting polygonal chain. For a simple polygonal chain, if the first vertex coincides with the last one $V_1=V_n$, i.e. their coordinates $(x_1,y_1)=(x_n,y_n)$, then it is a \textit{closed polygonal chain}; otherwise, it is an \textit{open polygonal chain}. In a simple polygon, two line segments meeting at a corner are usually required to form an angle that is not straight. However, we relax this constraint here and allow collinear line segments, since the polygonal chain analyzed in this paper is a sequence of discretization points representing the boundary of a tumor region in an image. Figure \ref{Figure 2}(a-c) show the example of an open, self-intersecting, and closed polygonal chain, respectively.

We mainly consider a closed polygonal chain in the paper, although this approach also works for a simple open polygonal chain with minor adjustments. The length of a closed polygonal chain is defined as the sum of the lengths of all line segments, i.e. $\sum_{i=1}^{n-1}\sqrt{(x_{i+1}-x_i)^2+(y_{i+1}-y_i)^2}$ since $V_1=V_n$. The center of a polygon is defined as the arithmetic mean position of all vertices, with its coordinate $(\sum_{i=1}^{n-1}x_i/(n-1),\sum_{i=1}^{n-1}y_i/(n-1))$. Without loss of generality, we assume that the polygonal chain $P$ has a unit length and a center in the space origin $(0,0)$. This can be done by rescaling and shifting each vertex $V_i$, i.e. transforming its coordinate,
\begin{equation*}
\begin{cases}
\begin{array}{l}
 \frac{1}{\sum_{i=1}^{n-1}\sqrt{(x_{i+1}-x_{i})^2+(y_{i+1}-y_{i})^2}}\left(x_i-\frac{1}{n-1}\sum_{i=1}^{n-1}x_i\right)\quad\mapsto\quad x_i\\
 \frac{1}{\sum_{i=1}^{n-1}\sqrt{(x_{i+1}-x_{i})^2+(y_{i+1}-y_{i})^2}}\left(y_i-\frac{1}{n-1}\sum_{i=1}^{n-1}y_i\right)\quad\mapsto\quad y_i
\end{array}
\end{cases}.
\end{equation*}

\subsection{Parameter Structure}\label{parameter} The proposed model maintains two evolving parameter groups. The first group, denoted by $\bm{\gamma}$, indicates the locations of all landmarks. The second group characterizes the deviation between the original polygon chain $P$ and the one formed by its landmarks.

\subsubsection{The landmark indicator vector}\label{landmark}
We define the \textit{landmarks} as those mathematically or structurally meaningful vertices in the boundary of a simple polygon, ignoring the remaining outline information. As the set of landmarks is a subset of $\{V_1,\ldots,V_{n-1}\}$, we use a latent binary vector $\bm{\gamma}=(\gamma_1,\ldots,\gamma_{n-1})$ to indicate which vertices are landmarks, with $\gamma_i=1$ if vertex $i$ is a landmark point and $\gamma_i=0$ otherwise for $i=1,\ldots,n-1$. Those landmarks constitute a reduced closed polygon chain namely the \textit{landmark chain} (see an example in Figure \ref{Figure 2}(d)), denoted by $P^{(\gamma)}$, where we use the superscript $(\gamma)$ to index the set of landmarks, characterized by $\gamma_j=1$. The number of vertices in $P^{(\gamma)}$ is the number of ones in $\bm{\gamma}$, denoted by $K=\sum_{i=1}^{n-1}\gamma_i$. The landmark chain can be also represented by a sequence of vertices $(V_{L_1},\ldots,V_{L_k},\ldots,V_{L_K},V_{L_1})$, where $L_k=\sum_{i=1}^{n-1}\delta\left(\sum_{j=1}^i\gamma_j=k\right)\delta(\gamma_i=1)$. Here, ${\delta}(\cdot)$ denotes the indicator function.

Those vertices whose $\gamma_j=0$ can be assigned into groups defined by the line segments bounded by two adjacent landmarks. Thus we use another vector $\bm{z}=(z_1,\ldots,z_{n-1}), z_i\in\{1,\ldots,K\}$ to reparameterize $\bm{\gamma}$, where $z_i=k$ if vertex $i$ is between landmarks $V_{L_k}$ and $V_{L_{(k+1)\text{ mod }K}}$, i.e. $V_i\in\{V_{L_k},\ldots,V_{L_{(k+1)\text{ mod }K}-1}\}$. Thus, we could consider the landmark identification as a segmentation problem (i.e. to partition $n-1$ vertices into $K$ mutually exclusive segments) or \textit{vice versa}. Mathematically, $\bm{z}$ is the cumulative sum of $\bm{\gamma}$, i.e. $z_i=\sum_{j=1}^i\gamma_j$ and all elements with zero-value are then filled with $K$, while $\bm{\gamma}$ is the lag one difference of $\bm{z}$, i.e. $\gamma_i=z_i-z_{i-1}$ and all negative elements are then filled with one. Note that $\bm{\gamma}$ and $\bm{z}$ are identical in the model in that both of them reveal the same information about the landmark locations. Figure \ref{Figure 2}(d) shows the representation of $\bm{\gamma}$ and $\bm{z}$ for an example of closed polygonal chain in Figure \ref{Figure 2}(c). Our goal is to find the landmark chain $P^{(\gamma)}$, i.e. to infer $\bm{\gamma}$ or $\bm{z}$, given a closed polygonal chain $P$ with coordinates $(x_1,y_1),\ldots,(x_{n-1},y_{n-1})$, subjecting to that $P^{(\gamma)}$ remains a closed polygonal chain.

To complete the model specification, we impose an independent Bernoulli (Bern) prior on $\bm{\gamma}$ as $\bm{\gamma} | \omega \sim  \prod_{i=1}^{n-1} \text{Bern}(\omega)$, where $\omega$ is interpreted as the probability of a vertex being a landmark \textit{a priori}. We further relax this assumption by allowing $\omega\sim\text{Beta}(\alpha_{\omega}, \beta_{\omega})$ to obtain a beta-Bernoulli prior on each $\gamma_j$. In practice, we suggest a constraint of $\alpha_{\omega} + \beta_{\omega} = 2$ for a vague prior setting \citep{tadesse2005bayesian}. Thus, if there are $\hat{K}$ landmarks expected \textit{a priori}, then we suggest to set $\alpha_{\omega}=2\hat{K}/n$ and $\beta_\omega=2(1-\hat{K}/n)$. In addition to that, we make the prior probability of $\bm{\gamma}$ equal to zero, if any of the following conditions are met: 1) there are less than three landmarks, i.e. $K<3$, in that a simple polygon should has at least three vertices; 2) two adjacent vertices are selected as landmarks, i.e. $\gamma_i=\gamma_{i+1}=1,\forall i$, in that a segment defined by $\bm{z}$ must contain at least two vertices; 3) $P^{(\gamma)}$ is a self-intersecting polygonal chain.

\subsubsection{The deviation between the polygonal and landmark chains}\label{deviation}
Here we discuss the probabilistic dependency between the observed closed polygonal chain $P$ and its landmark chain $P^{(\gamma)}$. We write the joint probability density function (p.d.f.) of $P$ as a product over the $K$ segments defined by its underlying landmarks, i.e. $f(V_1,\ldots,V_{n-1}|\bm{\gamma})=f(V_1,\ldots,V_{n-1}|\bm{z})=\prod_{k=1}^Kf(V_{L_k},\ldots,V_{L_{(k+1)\text{ mod }K}-1})$. A non-landmark vertex whose $\gamma_i=0$ should not be distant from the line segment defined by its two landmarks; otherwise, it might be considered as a landmark itself. Thus, we assume the shortest distance $d_i$ between vertex $V_i$ and the line segment in $P^{(\gamma)}$ that it corresponds to follows a distribution whose p.d.f. is a monotonically decreasing function with respect to the value of $d_i$. Suppose the straight line across the $k$ and $((k+1)\text{ mod }K)$-th landmarks, i.e. at locations $(x_{L_k},y_{L_k})$ and $(x_{L_{(k+1)\text{ mod }K}},y_{L_{(k+1)\text{ mod }K}})$, is expressed as $A_kx-B_ky+C_k=0$, where 
\begin{equation*}\label{line}
\begin{cases}
\begin{array}{l}
A_k\quad=\quad y_{L_{(k+1)\text{ mod }K}}-y_{L_k}\\
B_k\quad=\quad x_{L_{(k+1)\text{ mod }K}}-x_{L_k}\\
C_k\quad=\quad x_{L_{(k+1)\text{ mod }K}}  y_{L_k}-y_{L_{(k+1)\text{ mod }K}} x_{L_k}
\end{array}
\end{cases},
\end{equation*}
then the shortest distance between $V_i$ and the straight line across its corresponding landmarks can be computed as 
\begin{equation*}
d_i=\pm\frac{|A_kx_i-B_ky_i+C_k|}{\sqrt{A_k^2+B_k^2}}, 
\end{equation*}
where it is positive if $V_i$ is outside of the boundary of the landmark polygon $P^{(\gamma)}$ and negative otherwise.

We further assume that the shortest distances $d_i$'s of those non-landmark vertices between landmarks $k$ and $(k+1)\text{ mod }K$ are generated from a segment-specific zero-mean stationary Gaussian process (GP), modeling the spatial dependencies among local vertices through the covariance structure in a multivariate normal (MVN) distribution, that is,
\begin{equation}
f(V_{L_k},\ldots,V_{L_{(k+1)\text{ mod }K}-1}|\sigma_k^2)=\text{MVN}(d_{L_k+1},\ldots,d_{L_{(k+1)\text{ mod }K}-1};\bm{0},\sigma_k^2\bm{K}),
\end{equation}
where $\bm{0}$ is a $n_k$-by-$1$ all zero column vector, $\sigma_k^2$ is a scaling factor, and the kernel $\bm{K}$ is a $n_k$-by-$n_k$ positive definite matrix with each diagonal entry being one and each off-diagonal entry being a function of the relative position (e.g. Euclidean distance) between each pair of those non-landmark vertices. Note that $n_k=\sum_{i=1}^{n-1}\delta(z_i=k)-1$ is defined as the number of non-landmark vertices between landmarks $k$ and $(k+1)\text{ mod }K$. For the sake of simplicity, we choose to use the white noise kernel $\bm{K}=\bm{I}$, where $\bm{I}$ is a $n_k$-by-$n_k$ identity matrix, indicating that variability between each pair of $d_i$'s is uncorrelated. Note that this choice corresponds to a convex optimization problem of finding the best subset of $P$, that is $P^{(\gamma)}$, minimizing the $\ell_2$ norm of their deviations defined by the collection of shortest distances $\bm{d}=(d_1,\ldots,d_{n-1})$. Generalization of $\bm{K}$ to incorporate a certain spatial dependence structure or desired smoothness is left as future work.

Taking a conjugate Bayesian approach, we impose an inverse-gamma (IG) hyperprior on each $\sigma_k^2$, i.e., $\sigma_k^2\sim\text{IG}(\alpha_\sigma,\beta_\sigma)$. This parameterization setting is standard in most Bayesian univariate Gaussian models. It allows for creating a computationally efficient algorithm by integrating out the  variance component, which is usually a nuisance parameter. The integration leads to marginal non-standardized Student's $t$-distributions on $d_i$'s whose p.d.f. can be written as,
\begin{equation}\label{lklh}
\begin{aligned}
f\left(V_{L_k},\ldots,V_{L_{(k+1)\text{ mod }K}-1}\right)=&\int f\left(V_{L_k},\ldots,V_{L_{(k+1)\text{ mod }K}-1}|\sigma_k^2\right)\pi\left(\sigma_k^2\right){d}\sigma_k^2\\
=&(2\pi)^{-\frac{n_k}{2}}\frac{\Gamma\left(\alpha_\sigma+\frac{n_k}{2}\right)}{\Gamma(\alpha_\sigma)}\frac{\beta_\sigma^{\alpha_\sigma}}{\left(\beta_\sigma+\frac{1}{2}{\bm{d}_k^*}^T \bm{d}_k^*\right)^{\alpha_\sigma+\frac{n_k}{2}}},
\end{aligned}
\end{equation}
where $\bm{d}_k^*=(d_{L_k+1},\ldots,d_{L_{(k+1)\text{ mod }K}-1})$, i.e. the short distances of all non-landmark vertices assigned to segment $k$. To specify the IG hyperparameters, we recommend a choice by setting the shape parameter $\alpha_\phi$ to $3$, the minimum integer value such that the IG variance is defined, and the scale parameter $\beta_\phi$ to $1/(n-1)$, making both of the IG mean and standard deviation equal to ${1}/{(2n-2)}$. This choice is reasonable since we have already rescaled the polygonal chain $P$ to unit length. In the simulation study, we show that this choice is appropriate. 

\section{Model Fitting}\label{model fitting}
In this section, we describe the Markov chain Monte Carlo (MCMC) algorithms for posterior inference of the proposed method, that is, the inferential strategy to identify landmarks.

\subsection{MCMC Algorithm}\label{algorithm}
Our primary interest lies in the identification of landmarks via the inference on the landmark indicator vector $\bm{\gamma}$ given the polygonal chain $(V_1,\ldots,V_n)$ with the coordinates $(x_1,y_1),\ldots,(x_n,y_n)$. According to Section \ref{parameter}, the full data likelihood and the priors of the landmark detection model can be written as
\begin{equation}\label{full_lklh}
\begin{aligned}
f(V_1,\ldots,V_{n-1}|\bm{\gamma})=&\prod_{k=1}^Kf\left(V_{L_k},\ldots,V_{L_{(k+1)\text{ mod }K}-1}\right)\\
=&(2\pi)^{-\frac{n-K}{2}}\left(\frac{\beta_\sigma^{\alpha_\sigma}}{\Gamma(\alpha_\sigma)}\right)^K\prod_{k=1}^K\frac{\Gamma\left(\alpha_\sigma+\frac{n_k}{2}\right)}{\left(\beta_\sigma+\frac{1}{2}{\bm{d}_k^*}^T \bm{d}_k^*\right)^{\alpha_\sigma+\frac{n_k}{2}}}\text{ and }\\
\pi(\bm{\gamma})=&\frac{\Gamma(\alpha_\omega+\beta_\omega)}{\Gamma(\alpha_\omega)\Gamma(\beta_\omega)}\frac{\Gamma(\alpha_\omega+K)\Gamma(\beta_\omega+n-K)}{\Gamma(\alpha_\omega+\beta_\omega+n)},
\end{aligned}
\end{equation}
respectively. To serve this purpose via sampling from the posterior distribution $\pi(\bm{\gamma}|V_1,\ldots,V_{n-1})\propto f(V_1,\ldots,V_{n-1}|\bm{\gamma})\pi(\bm{\gamma})$, an MCMC algorithm is designed based on Metropolis search variable selection algorithms \citep{george1997approaches,brown1998multivariate}. As aforementioned in Section \ref{deviation}, we have integrated out the variance components $\sigma_1^2,\ldots,\sigma_K^2$. This step helps us speed up the MCMC convergence and improve the estimation of $\bm{\gamma}$. The factorization in the full data likelihood allows Hastings ratios to be evaluated locally with respect to the affected segments. Note that this algorithm is sufficient to guarantee ergodicity for our model. 

The BayesLASA proceeds through iterations after initialization, each of which updates the configuration of $\bm{\gamma}$. Within each iteration $b$, a new candidate $\bm{\gamma}^*$ is generated via an \textit{add-delete-swap} algorithm. The proposed move is accepted, $\bm{\gamma}^{(b)}=\bm{\gamma}^*$, with probability $\min(1,m)$; otherwise, the move is rejected, $\bm{\gamma}^{(b)}=\bm{\gamma}^{(b-1)}$. The Hastings ratio $m$ is computed as,
\begin{equation*}
m=\frac{\pi\left(\bm{\gamma}^*|V_1,\ldots,V_{n-1}\right)}{\pi\left(\bm{\gamma}^{(b-1)}|V_1,\ldots,V_{n-1}\right)}\frac{J\left(\bm{\gamma}^{(b-1)}|\bm{\gamma}^*\right)}{J\left(\bm{\gamma}^*|\bm{\gamma}^{(b-1)}\right)},
\end{equation*}
where $J\left(\bm{\gamma}^*|\bm{\gamma}^{(b-1)}\right)$ is the proposal density, which specifies the probability of proposing a move to $\bm{\gamma}^*$ given the previous state $\bm{\gamma}^{(b-1)}$, and $J\left(\bm{\gamma}^{(b-1)}|\bm{\gamma}^*\right)$ is the flipped case.

For the \textit{add-delete} step, a new candidate vector, say $\bm{\gamma}^*$, is generated by randomly choosing an entry within $\bm{\gamma}$, say the $i$-th one, and changing its value to $1-\gamma_i$. If $\gamma_i=0$, then $\gamma_i^*=1$, indicating that the $k$-th segment in $\bm{z}$, which vertex $i$ belongs to, has been split into two segments composed of $(V_{L_k},\ldots,V_{i-1})$ and $(V_i,\ldots,V_{L_{(k+1)\text{ mod }K}-1})$, respectively. If $\gamma_i=1$, then $\gamma_i^*=0$, indicating that the $k-1$ and $k$-th segment in $\bm{z}$, where vertex $i$ is their common endpoint, have been merged into one segment formed by $(V_{L_{k-1}},\ldots,V_{L_{(k+1)\text{ mod }K}-1})$. For the \textit{swap} step, we randomly choose an non-landmark vertex in $P$, say the $i$-th vertex, and a landmark, say the $i'$-th vertex. Then, we change the value of $\gamma_i$ from $0$ to $1$, while setting the value of $\gamma_{i'}$ from $1$ to $0$. This corresponds to splitting a segment into two segments and merging two adjacent segments into one segment in $\bm{z}$ simultaneously. In addition to that, we propose another type of move namely \textit{partial shift} suggested by \cite{li2011mcmc} to make the Markov chain fast convergent. Specifically, we shift the sequence $\bm{\gamma}$ left or right by up to a pre-specified number, say $s=\pm1$, where the negative number indicates left shifts and the positive number means right shifts. We set each entry $\gamma_i^*$ with the value of $\gamma_{(i+s)\text{ mod } (n-1)}$. To improve mixing, we suggest to perform the last two types of moves in every $20$ iteration. In the applications of this paper, no improvement in the MCMC performance was noticed beyond this value. 

\subsection{Posterior Estimation}
We explore posterior inference on the landmark locations $\bm{\gamma}$ by postprocessing the MCMC samples after burn-in, denoted by $(\bm{\gamma}^{(1)},\ldots,\bm{\gamma}^{(B)})$, where $b$ indexes the MCMC iteration after burn-in from now on. One way is to choose the $\bm{\gamma}$ corresponding to the \textit{maximum-a-posteriori} (MAP),
\begin{equation*}
\hat{\bm{\gamma}}^{\text{MAP}}=\underset{b}{\text{argmax}}~\pi\left(\bm{\gamma}^{(b)}|V_1,\ldots,V_{n-1}\right).
\end{equation*}
The corresponding $\hat{\bm{z}}^{\text{MAP}}$ can be obtained by taking the cumulative sum of $\hat{\bm{\gamma}}^{\text{MAP}}$. An alternative estimate relies on the computation of posterior pairwise probability matrix (PPM), that is, the probabilities that vertices $i$ and $i'$ are assigned into the same segment in $\bm{z}$, $c_{ii'}=\sum_{b=1}^B\delta(z_i^{(b)}=z_{i'}^{(b)}|V_1,\ldots,V_{n-1})$. This estimate, suggested by \cite{dahl2006model}, utilizes the information from all MCMC samples and is thus more robust. After obtaining this $(n-1)$-by-$(n-1)$ co-clustering matrix denoted by $\bm{C}$, a point estimate of $\bm{z}$ can be approximated by minimizing the sum of squared deviations of its association matrix from the PPM, that is,
\begin{equation*}
\hat{\bm{z}}^{\text{PPM}} = \underset{\bm{z}}{\text{argmin}}~ \sum_{i < i'} \left( \delta(z_i = z_{i'}) - c_{ii'} \right)^2.
\end{equation*}
The corresponding $\hat{\bm{\gamma}}^{\text{PPM}}$ can be obtained by taking the difference between consecutive entries in $\hat{\bm{z}}^{\text{PPM}}$. 

To construct a ``credible interval" for a change point, we utilize its local dependency structure from all MCMC samples of $\bm{\gamma}$ that belong to its neighbors. As aforementioned in Section \ref{landmark}, if vertex $i$ is selected as a landmark, then its adjacent vertices $i\pm1$ must not be a landmark, because the prior probability of $\bm{\gamma}$ is set to be zero in this case. Thus the correlation between the MCMC sample vectors $(\gamma_t^{(1)},\ldots,\gamma_t^{(B)})$ and $(\gamma_{t\pm u}^{(1)},\ldots,\gamma_{t\pm u}^{(B)})$ tends to be negative when $u$ is small. Thus, we define the credible interval of a landmark as the two ends of all its nearby consecutive vertices, of which MCMC samples of $\bm{\gamma}$ are significantly negatively correlated with that of the landmark. This could be done via a one-sided Pearson correlation test with a pre-specified significant level, e.g. $0.05$. 

\section{Simulation}\label{simulation}
In this section, we use simulated data generated from the proposed model to assess our inferential strategy's performance for landmark detection against alternative solutions. It is shown that the BayesLASA outperforms other approaches in terms of both landmark identification accuracy and computational efficiency.  

Simulated data were generated in the following steps. We first randomly generated an equilateral or non-equilateral simple polygon with $K=4$, $5$, or $6$ points (considered as landmarks) in a Cartesian plane, corresponding to a quadrilateral, pentagon, or hexagon, respectively. The lengths of edges that make up the simple polygon uniformly ranged from $50$ to $100$. Next, we ``binned" the landmark chain into a series of $n=140$, $150$, $160$, $170$, or $180$ equally sized intervals. Then, for each underlying interval, a vertex was generated with its perpendicular distance to the interval sampled from $\text{N}(0,\sigma^2)$, where the scaling factor $\sigma^2$ was chosen from $\{0.5,2,4\}$. We sequentially connected all vertices, including the $K$ landmarks, to form a closed polygonal chain. Last, we rescaled and shifted the generated polygonal chain so that its length is one and its centroid is at $(0,0)$. Note that the starting point of the chain was selected at random. Since we had three, four, and three choices of $K$, $n$, and $\sigma^2$, respectively, there were $3\times5\times3=45$ scenarios in total. For each scenario, we repeated the above steps to generate $50$ independent datasets.

As for the Beta prior on the landmark selection parameter $\omega$, we set the two hyperparameters $\alpha_\omega=2\hat{K}/n$ and $\beta_\omega=2(1-\hat{K}/n)$, where $\hat{K}=3$, corresponding to three landmarks or a proportion of $3/n$ vertices being landmarks expected \textit{a priori}. As for the IG prior on the non-landmark characteristic parameter $\sigma_k^2$, we set the two hyperparameter $\alpha_\sigma=3$ and $\beta_\sigma=1/n$ as discussed in Section \ref{deviation}. As for the BayesLASA's algorithm setting implemented in this paper, we ran the MCMC chain with $100n$ iterations, discarding the first $50\%$ sweeps as burn-in. We started the chain from a model by randomly assigning $\hat{K}=3$ ones in the landmark indicator vector $\bm{\gamma}$, while all other entries were zeros.

To evaluate the landmark identification accuracy, we might rely on the binary landmark indicator vector $\bm{\gamma}$. Since landmarks and non-landmark vertices are usually of very different sizes (i.e. landmarks are usually assumed to be a small fraction of all vertices), most of the binary classification metrics are not suitable for model comparison. Thus, we chose to use the Matthews correction coefficient (MCC) \citep{matthews1975comparison}, which is defined as \[\text{MCC}(\bm{\gamma}, \hat{\bm{\gamma}})=\frac{(\text{TP}\times \text{TN} - \text{FP}\times \text{FN})}{\sqrt{(\text{TP+FP})(\text{TP+FN})(\text{TN+FP})(\text{TN+FN})}},\] 
where TP, TN, FP, and FN stand for true positive, true negative, false positive, and false negative. They are the four entries in the confusion matrix that can be summarized from the estimated $\hat{\bm{\gamma}}$ and true $\bm{\gamma}$. The MCC value ranges from $-1$ to $1$. The larger the MCC, the more accurate the inference. In addition, we evaluated the model performance through the segmentation vector $\bm{z}$ based on the adjusted Rand index (ARI) \citep{hubert1985comparing}. The ARI is the corrected-for-chance version of the Rand index \citep{rand1971objective}, as a similarity measure between two sample allocation vectors. Let $N_1=\sum_{i > i'}{\delta}(z_i = z_{i'})\text{}(\hat{z}_{i} = \hat{z}_{i'})$, $N_2=\sum_{i > i'}{\delta}(z_i = z_{i'}) {\delta}(\hat{z}_{i} \neq \hat{z}_{i'})$, $N_3=\sum_{i > i'}{\delta}(z_i \neq z_{i'}) {\delta}(\hat{z}_{i} = \hat{z}_{i'})$, and $N_4=\sum_{i > i'}{\delta}(z_i \neq z_{i'}) {\delta}(\hat{z}_{i} \neq \hat{z}_{i'})$ be the number of pairs of vertices that are 1) in the same segment in both of the true and estimated partitions; 2) in different segments in the true partition but in the same segment of the estimated one; 3) in the same segment of the true partition but in different segments in the estimated one; and 4) in different segments in both of the true and estimated partitions. Then, the ARI can be computed as
\[\text{ARI}(\bm{z}, \hat{\bm{z}})=\frac{\binom{n-1}{2}(N_1 + N_4) - [(N_1 + N_2)(N_1 + N_3) +(N_3+ N_4)(N_2+N_4)]}{\binom{n-1}{2}^2 - 
	[(N_1+N_2)(N_1+N_3) +(N_3+N_4)(N_2+N_4)]}.\]
The ARI usually yields values between $0$ and $1$, although it can yield negative values \citep{santos2009use}. The large the ARI, the more similar between $\bm{z}$ and $\hat{\bm{z}}$, thus the more accurately the method detects the true landmarks. 

There is no method like the BayesLASA that can detect landmarks for a single polygonal chain to the best of our knowledge. In setting up a comparison study, we considered a recently developed algorithm named automatic landmark detection model for planar shape data (ALDUQ) \citep{strait2019automatic} based on Bayesian inference. It detects the number and locations of landmarks using square-root velocity function representation under the elastic curve paradigm. Thus, we needed to convert the discrete polygonal chain into a continuously differentiable curve using the Gaussian kernel smoother with an appropriate length scale parameter (e.g. $0.05$). The output of ALDUQ included the relative locations of landmarks and their credible intervals presented by arc-lengths. To make a feasible comparison, we considered a landmark correctly identified if its estimated location was within a local window with length $5$ of the true position. As for the ALDUQ's algorithm setting, we used the same MCMC iteration numbers $100n$. The vertices of the convex hull of a simple polygon were also considered to be landmarks in practice. Thus, we evaluated the performance when $\hat{\gamma}$ were those vertices as a benchmark.

\begin{figure}[h]
	\begin{center}
		\includegraphics[width=1.0\linewidth]{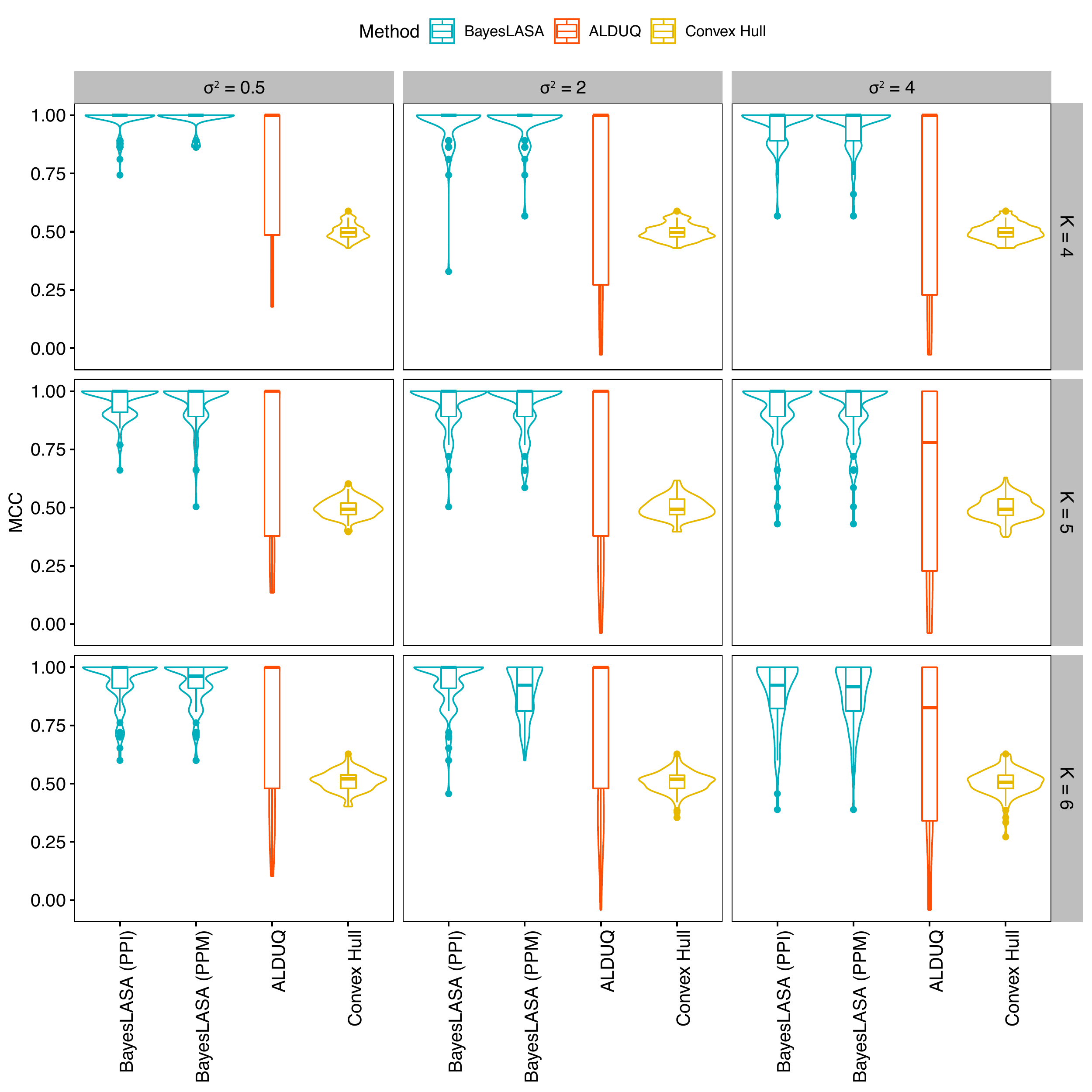}
	\end{center}
	\caption{Simulation study: The violin plots of Matthews correction coefficients (MCCs) from replicated datasets generated under different scenarios in terms of true landmark number $K$ and noise level $\sigma^2$. Blue, red, and yellow violins correspond to the results obtained by the BayesLASA, ALDUQ \citep{strait2019automatic}, and benchmark convex hull.}
	\label{Figure 3a}
\end{figure}

\begin{figure}[h]
	\begin{center}
		\includegraphics[width=1.0\linewidth]{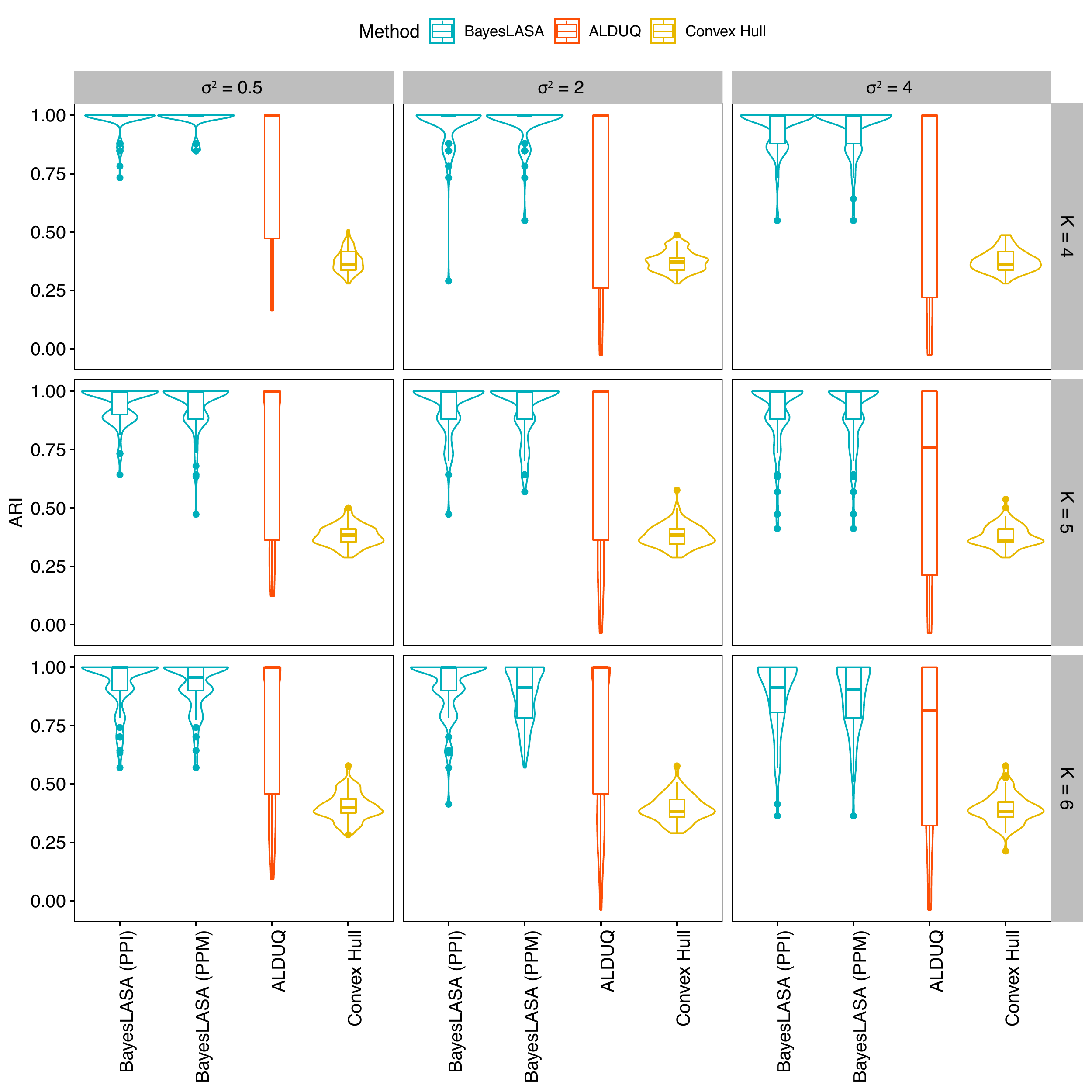}
	\end{center}
	\caption{Simulation study: The violin plots of adjusted Rand indexes (ARIs) from replicated datasets generated under different scenarios in terms of true landmark number $K$ and noise level $\sigma^2$. Blue, red, and yellow violins correspond to the results obtained by the BayesLASA, ALDUQ \citep{strait2019automatic}, and benchmark convex hull.}
	\label{Figure 3b}
\end{figure}

Figure \ref{Figure 3a} and \ref{Figure 3b} exhibit the landmark detection performances for each scenario, while those with the different $n$ were grouped. Our BayesLASA performed much better than the ALDUQ and benchmark convex hull in terms of both MCC and ARI. We also noted that all three approaches' accuracy decreased as the number of true landmarks $K$ or the scaling factor $\sigma^2$ increased, leading to a more complex polygonal chain.
\begin{figure}[h]
	\begin{center}
		\includegraphics[width=1.0\linewidth]{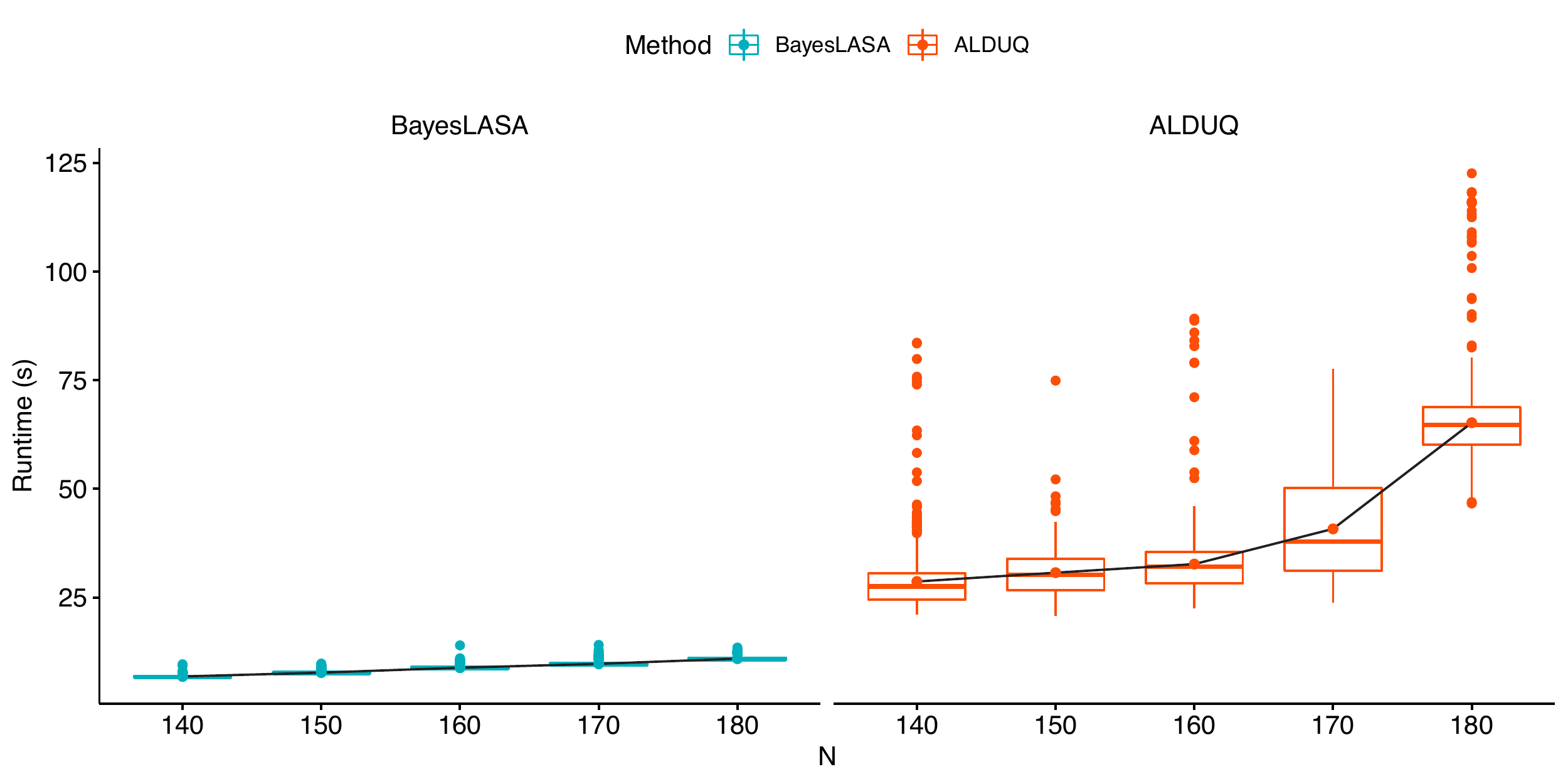}
	\end{center}
	\caption{Simulation study: The box plots of runtime in seconds from replicated datasets generated under different scenarios in terms of vertex number $n$. Blue and red boxes correspond to the results obtained by the BayesLASA and ALDUQ \citep{strait2019automatic}.}
	\label{Figure 3c}
\end{figure}

We conclude the section by conducting a scalability test in \texttt{R} with \texttt{Rcpp} package \citep{eddelbuettel2011rcpp} on a Mac PC with $2.30$GHz CPU and $8$ GB memory. Our simulated datasets were of five cases of $n$, i.e. $n=\{140, 150, 160, 170, 180\}$. Both the BayesLASA and ALDUQ were run with $100n$ MCMC iterations. Figure~\ref{Figure 3c} shows the runtime as a function of $n$. Our approach outperformed the ALDUQ obviously in terms of time efficiency. We further fit a linear regression for the runtime against $n$ and obtained an $R^2$ of $0.929$, suggesting that the actual runtime increased approximately linearly in the number of vertices in a polygonal chain. The estimated model was $\text{runtime} = -160.2 + 343.5n$. This analysis implied the proposed BayesLASA can be applied to a polygonal chain with a large number of vertices $n$.

\section{Case Study on Lung Cancer}\label{application}
Lung cancer has been ranked as the leading cause of death from cancer, with non-small-cell lung cancer (NSCLC) accounting for about $85\%$ of lung cancer deaths. Current guidelines for diagnosing and treating cancer are largely based on pathological examination of tissue section slides. A deep-learning approach has been developed to perform the tumor segmentation of pathology images in our previous studies \citep{wang2018comprehensive}. Specifically, a convolutional neural network (CNN)-based classifier was trained using a large cohort of lung cancer pathology images manually labeled by well-experienced pathologists. It can classify each image patch into one of the three categories: normal, tumor, or white (background). In this case study, we used $246$ pathology images from $143$ NSCLC patients in the National Lung Screening Trial (NLST). Each patient has one or more tissue slide(s) scanned at $40\times$ magnification. The median size of the slides is $24,244\times19,261$ pixel. Segmentation was done by the CNN classifier and the boundary of each tumor region was then extracted and presented as a simple closed polygonal chain. The tumor region with the largest area in each slide was considered in our following analysis.

\begin{figure}
	\centering
	\includegraphics[width= 1 \textwidth]{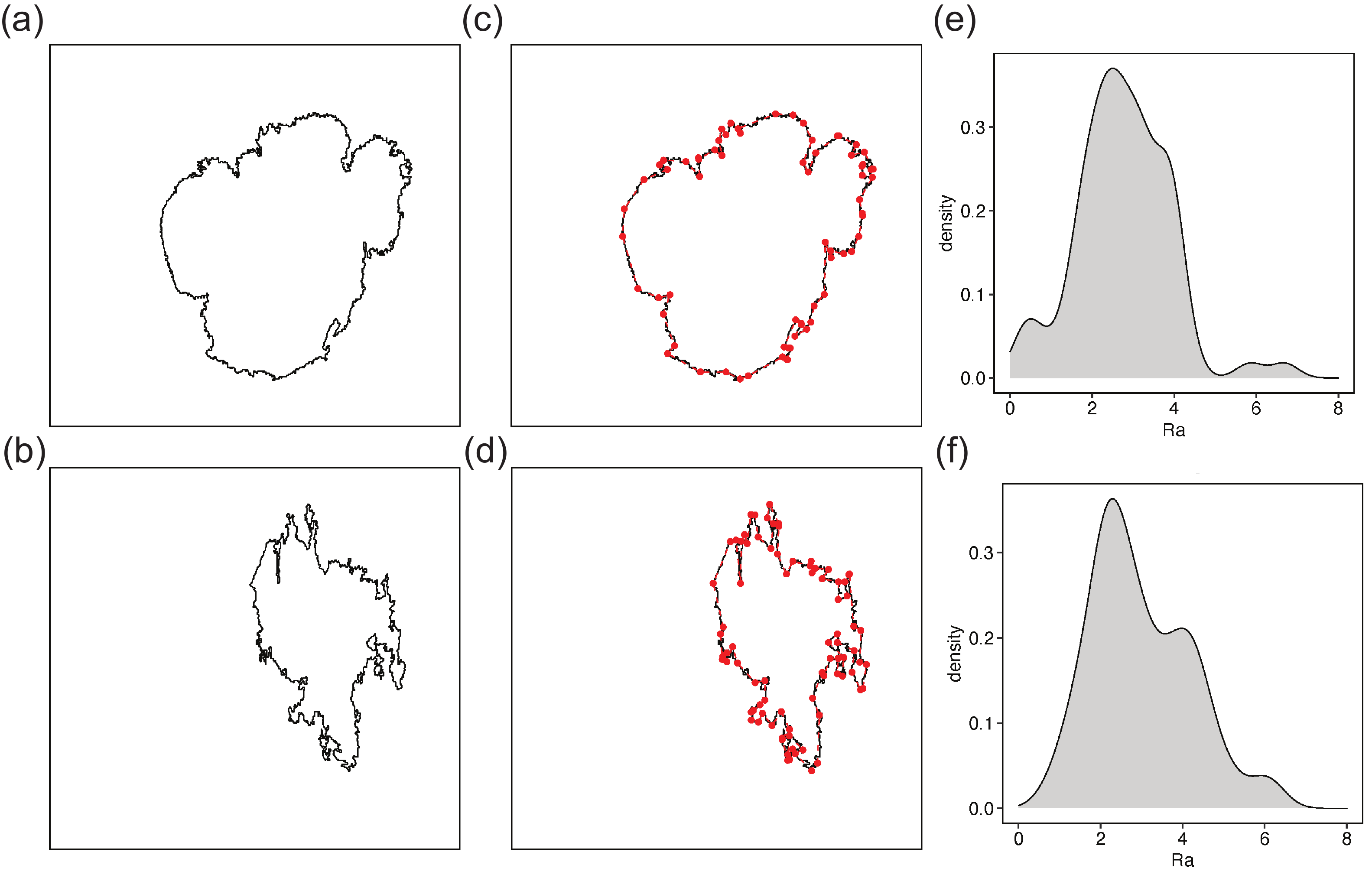}
	\caption{Lung cancer case study: (a) and (b) Two examples of the extracted tumor boundaries in the National Lung Screening Trial (NLST) dataset; (c) and (d) The landmarks (shown in red) identified by the proposed BayesLASA for the tumor boundaries as shown in (a) and (b); (e) and (f): The density plots of piecewise roughness measurements Ra determined by the landmarks as shown in (c) and (d).}
	\label{f:fig_eg}
\end{figure}

We applied the proposed model with the same hyperparameter and algorithm settings as described in Section \ref{simulation} except $\beta_{\sigma} = 500$. A total of four MCMC chains were run simultaneously and averaged PPM estimates on $\bm{\gamma}$ was used to determine the landmarks for each tumor. Figure~\ref{f:fig_eg} shows two example of tumor boundaries and their identified landmarks by the BayesLASA from a patient with good prognosis and another patient with poor prognosis. Notably, tumor from the patient with shorter survival time exhibited a more spiculated shape compare to the one with good prognosis, indicating the invasion of tumor cells into surrounding tissues. Although the two tumor regions have distinctive tumor boundaries, the roughness and its heterogeneity are much more subtle for many other examples. Therefore, the proposed BayesLASA can be used to predict the survival time when human visualization does not work.  

The landmark chain $P^{(\gamma)}$ was used as a skeleton reference and for each segment in $\bm{z}$. The distance between the original polygonal chain $P$ and the latent landmark chain $P^{(\gamma)}$ can be presented as $\bm{d}=(d_1,\ldots,d_{n-1})$. Two types of tumor boundary roughness features, distanced and model-based were employed based on the segmentation $\bm{z}$ and distances $\bm{d}$.

\textbf{Distance-based features:}
Surface roughness measurements computed by simple math equations were adopted to quantify the irregularity of tumor boundary. Definition of those eight measurements were summarized in Table~\ref{t:tab_r}. For each pathology image slide, surface roughness measurements were computed for each segment.
\begin{table}[htbp]
	\caption{A list of distance-based surface roughness measurements considered in this paper.}
	\label{t:tab_r}
	\renewcommand{\arraystretch}{1.2}
	\begin{center}
		\begin{tabular}{lll}
			\hline
			& {Description} & {Definition} \\
			\hline
			Ra    & Arithmetical mean deviation & $\sum_{i=1}^n |d_i|/n$ \\
			Rq    & Root mean squared & $\sqrt{\sum_{i=1}^n d_i^2/n}$ \\
			Rv    & Maximum valley depth &  $|\text{min} (d_i) |$ \\
			Rp    & Maximum peak height & $\text{max}(d_i)$ \\
			Rz    & Maximum height of the profile &  Rv+Rp \\
			Rsk   & Skewness & $\sum d_i^3/(n\text{Rq}^3)$ \\
			Rku   & Kurtosis & $\sum d_i^4/(n\text{Rq}^4)$ \\
			RzJIS & Based on the five highest peaks and lowest valleys  & $\sum_{i=1}^5 (\text{Rp}_i - \text{Rv}_i)/5$ \\
			\hline
		\end{tabular}%
	\end{center}
\end{table}

\textbf{Model-based features:}
Since the distances $d_i$'s are sequentially indexed, their changing frequencies indicated the fluctuation degree of surface roughness. A hidden Markov model (HMM) with Gaussian emission was employed to fit $\bm{d}$. The observed values of $d_i$'s were from a two-component Gaussian mixtures. Two hidden states, corresponding to the two components, were defined to illustrate the negative `$-$' (within the landmark chain) and positive `$+$' (out of the landmark chain) sign of each entry $d_i$. The transition probabilities control the way that the hidden state of $d_{i+1}$ is chosen given the hidden state of $d_i$ within each segment determined by $\bm{z}$, reflecting the segment-specified roughness. The transition probabilities, denoted by $a_{++}$, $a_{+-}$, $a_{-+}$, and $a_{--}$, respectively, were estimated for each segment.

\subsection{Association study}
With the identified landmarks for all tumor regions, we conducted a downstream analysis to investigate their associations with other interest measurements, such as clinical outcomes. Specifically, a multivariate Cox proportional hazard (CoxPH) model \citep{Cox1992} was fitted with the summary statistics such as the mean, standard deviation, skewness, and kurtosis of the piecewise roughness measurements, after adjusting for the number of landmarks $K$, tumor size and other clinical information, such as gender, tobacco history, and cancer stage. Multiple sample images from the same patient were modeled as correlated observations in the CoxPH model to compute a robust variance for each coefficient. 

The outputs of the CoxPH model using one of the distance-based features, Ra, is shown in Table~\ref{t:ra_cox}. More advanced stage of lung cancer is significantly correlated with poorer prognosis, which is in consonance with previous knowledge. It is noteworthy that significant effects of kurtosis ($\text{Coef}<0$) and skewness ($\text{Coef}>0$). The same result were observed for the other choices of distance-based surface roughness measurements including Rq, Rp, Rv, Rx, and RzJIS (shown in the Table S1 - S5 in the supplement). CoxPH model fitted with Ra obtained an overall $p$-value of $0.0001$ (Wald test). The CoxPH model fitted with the moments of Ra suggested the roles as prognostic factors of kurtosis and skewness (both $p$-values $< 0.001$), which measure the peakedness and asymmetry of the probability distribution respectively. The negative coefficient of kurtosis and positive coefficient of skewness suggested tumor with smaller kurtosis (flat spreading) and larger skewness (left-centered) are more heterogeneous in surface roughness and therefore indicate the worse prognosis of patients (as illustrated in Figure~\ref{f:fig_eg}(e) and (f)). It is consistent with the biology knowledge that high spatial heterogeneity is a pivotal feature of cancer at a cellular and histological level resulted from the distinct patterns of different cancer cell subpopulations in terms of dysregulation of proliferation, mobility, and metabolism pathways \citep{meacham2013tumour,dagogo2018tumour}. The underlying biological mechanism of heterogeneous tumor boundary could attribute to the heterogeneous regulation of gene expression by abnormally activated Rho GTPases pathways among cancer cell subpopulations and the consequent dissimilarity in the  downstream actin cytoskeleton and stress fibers \citep{pascual2017rnai}.  
\begin{table}
	\caption{Lung cancer case study: The outputs of fitting a multivariate Cox proportional hazards (CoxPH) model with survival time and vital status as responses and summary statistics of distance-based piecewise roughness measurements, i.e. Ra's, tumor size, cancer stage, tobacco history, and gender as predictors.}
	\label{t:ra_cox}
	\renewcommand{\arraystretch}{1.2}
	\begin{center}
		\begin{tabular}{lrccl}
			\hline
			{Predictor} & {Coef} & exp(Coef) & {SE} & {$p$-value}\\
			\hline
			Mean & $-0.757$ & $0.469$ & $0.604$  & $0.210$\\
			Standard deviation & $-0.428$ & $0.652$ & $0.627$ & $0.494$\\
			Kurtosis & $-0.368$ & $0.692$ & $0.106$ & $ \bm{5.4\times10^{-4}}$\\
			Skewness & $1.675$ & $5.336$ & $0.469$  & $\bm{1.8\times10^{-4}}$\\
			Number of landmarks $K$ & $0.014$ & $1.014$ & $0.007$& $\bm{0.047}$\\
			Area & $0.000$ & $1.000$ &$0.000$ & $0.828$\\
			Cancer stage II vs. I & $0.343$ & $1.410$ & $0.619$  & $0.579$\\
			Cancer stage III vs. I & $1.168$ & $3.214$ &$0.371$  & $\bm{0.002}$\\
			Cancer stage IV vs. I & $1.802$ & $6.064$ & $0.463$ & $\bm{9.9\times10^{-5}}$\\
			Smoking vs. non-smoking & $-0.119$ & $0.888$ & $0.330$  & $0.718$\\
			Female vs. male & $-0.127$ & $0.881$ & $0.322$ & $0.694$\\
			\hline
			\multicolumn{5}{p{0.8\textwidth}}{Abbreviations: Coef is coefficient and SE is standard error.}
		\end{tabular}
	\end{center}
\end{table}

To validate that our landmark-based shape analysis is robust to the choice of roughness measurements, we repeated the above step to fit another CoxPH model with the model-based roughness measurements as the predictors. The overall $p$-value of CoxPH model fitted with moments of negative to positive transition probability $a_{-+}$ is $0.0008$ (Wald test) and the coefficients and $p$-values for each variable are shown in Table~\ref{t:neg2pos_cox}. The CoxPH model for the opposed transition probability $a_{+-}$ was summarized in Table S6 in the supplement, showing a similar result. Again, kurtosis and skewness are significant factors associated with patient outcomes. Furthermore, standard deviation had a $p$-value $=0.0009$ and a large positive coefficient, implying that the larger heterogeneous the tumor boundary roughness was, the poorer prognosis the patient had. 
\begin{table}
	\caption{Lung cancer case study: The outputs of fitting a multivariate Cox proportional hazards (CoxPH) model with survival time and vital status as responses and summary statistics of model-based piecewise roughness measurements, i.e. hidden Markov model (HMM) transition probabilities $a_{-+}$, tumor size, cancer stage, tobacco history, and gender as predictors.}
	\label{t:neg2pos_cox}
	\renewcommand{\arraystretch}{1.2}
	\begin{center}
		\begin{tabular}{lrccl}
			\hline
			{Predictor} & {Coef} & exp(Coef) & {SE} & {$p$-value}\\\hline
			Mean & $5.797$ & $329.4$ & $9.541$ & $0.543$\\
			Standard deviation & $18.10$ & $7.3\times10^{7}$ & $6.915$ & $\bm{0.009}$\\
			Kurtosis & $0.091$ & $1.096$ & $0.046$ & $\bm{0.046}$\\
			Skewness & $-0.958$ & $0.384$ & $0.369$ & $\bm{0.009}$\\
			Number of landmarks $K$ & $0.014$ & $1.014$ & $0.008$ & $0.088$\\
			Area & $0.000$ & $1.000$ & $0.000$ & $0.783$\\
			Cancer stage II vs. I & $0.502$ & $1.651$ & $0.593$ & $0.397$\\
			Cancer stage III vs. I & $1.195$ & $3.303$ & $0.393$ & $\bm{0.002}$\\
			Cancer stage IV vs. I & $1.791$ & $5.997$ & $0.490$ & $\bm{2.6\times10^{-4}}$\\
			Smoking vs. non-smoking & $-0.042$ & $0.959$ & $0.322$ & $0.896$\\
			Female vs. male & $-0.122$ & $0.885$ & $0.307$ & $0.692$\\
			\hline
			\multicolumn{5}{p{0.8\textwidth}}{Abbreviations: Coef is coefficient and SE is standard error.}
		\end{tabular}
	\end{center}
\end{table}

\begin{figure}
	\centering
	\includegraphics[width= 0.6 \textwidth]{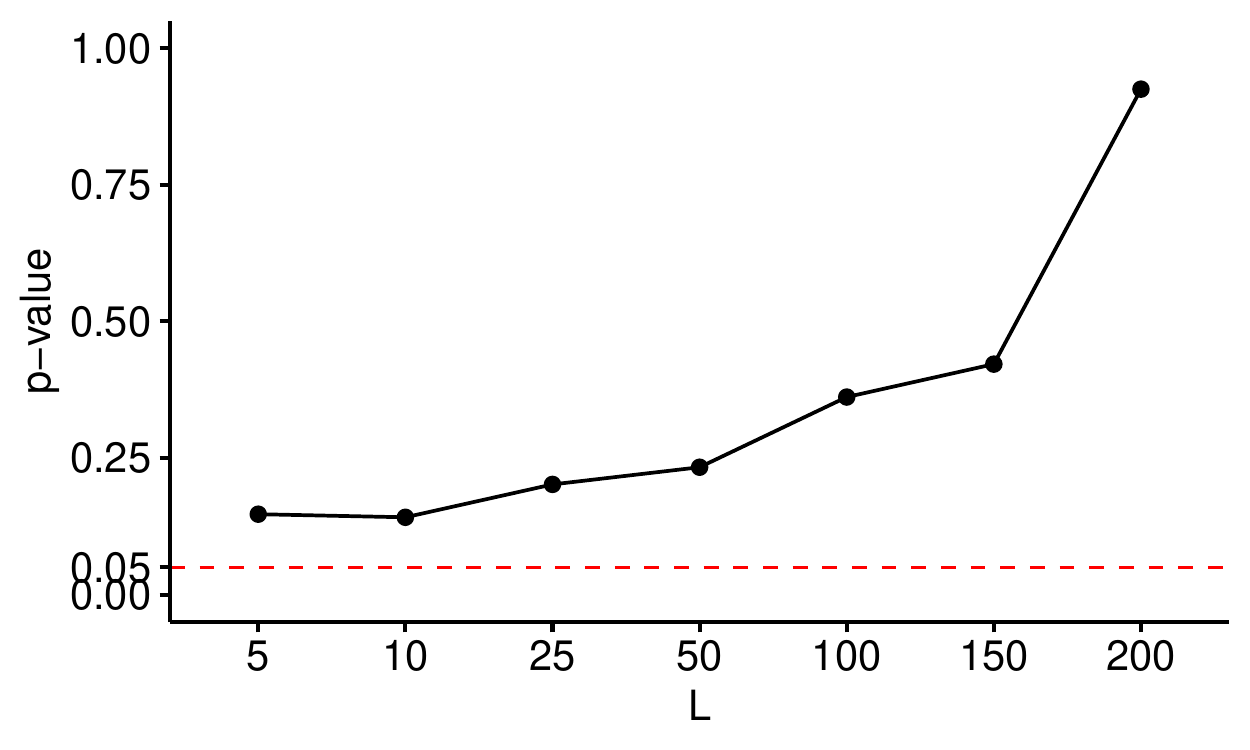}
	\caption{Lung cancer case study: The $p$-values, ranging from $0.142$ to $0.925$, of the tumor boundary roughness (TBR) under different choices of window size $L$ by fitting a multivariate Cox proportional hazards (CoxPH) model with survival time and vital status as responses and TBR, tumor size, cancer stage, tobacco history, and gender as predictors. The red dash line indicates a significant level of $0.05$.}
	\label{f:fig_tbr_l}
\end{figure}

By contrast, we fitted a similar CoxPH model by using the radial distance-based shape features as predictors, including the zero-crossing count (ZCC) and tumor boundary roughness (TBR). Let a simple closed polygonal chain $(V_1,\ldots,V_{n-1})$ denote the tumor boundary pixels in a medical image, with the coordinate of vertex $i$ being $(x_i,y_i)$ for $i=1,\ldots,n-1$. Then the radial distance $r_i$ between vertex $i$ and the polygon center, of which coordinate is $(\bar{x},\bar{y})=(\sum_{i=1}^{n=1}x_i/(n-1),\sum_{i=1}^{n-1}y_i/(n-1))$, is defined as
\[r_i=\sqrt{(x_i-\bar{x})^2+(y_i-\bar{y})^2}.\]
By tracing the radial distances of all vertices, i.e. $\bm{r}=(r_1,\ldots,r_{n-1})$, the ZCC is the count number  of times radial length crosses the mean value $\bar{r}=\sum_{i=1}^{n-1}r_i/(n-1)$,
\[\text{ZCC}=\sum_{i=1}^{n-1}\delta\left((r_i-\bar{r})(r_{i+1}-\bar{r})<0\right).\]
The TBR is calculated by averaging the roughness index (RI) for a window with length $L$ over the entire tumor boundary, where the RI for window $j$ is defined as $\text{RI}_j=\sum_{i=(j-1)L+1}^{jL-1} |r_{i+1}-r_{i}|$ for $j=1,\ldots,\ceil{{(n-1)}/{L}}$,
\[\text{TBR}=\sum_{j=1}^{\ceil{{(n-1)}/{L}}}\text{RI}_j/\ceil{{(n-1)}/{L}}.\]
Here $\lceil\cdot\rceil$ denotes the ceiling function. The results implied an insignificant association between the ZCC and clinical outcomes ($p$-value = $0.128$). As for the TBR, we tried to vary the window size $L$ from $5$ to $200$. Figure~\ref{f:fig_tbr_l} shows the TBR $p$-values against $L$. The obtained $p$-values ranged from $0.142$ to $0.925$. Unfortunately, we could not find any association between these two radial distance-based roughness measurements and the patient survival outcome from the NLST dataset. The comparison demonstrates that the proposed model-based shape analysis can lead to sharper inferences on tumor boundary roughness than ordinary exploratory analyses.

\subsubsection{Predictive Performance by Cross-Validation}\label{s:loov_ra}
\begin{figure}[!h]
	\centering
	\subfloat[Distance-based roughness measurements]{%
		\includegraphics[width=0.49\columnwidth]{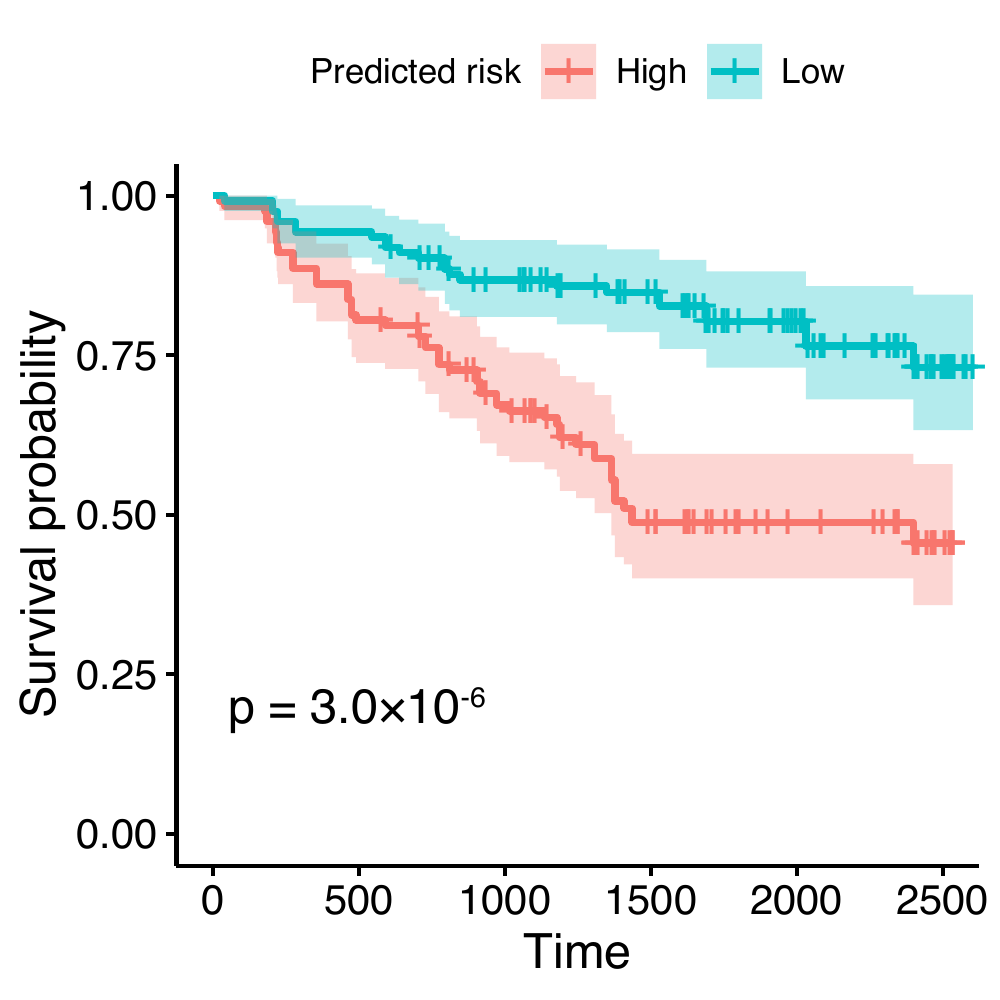}
	}
	\subfloat[Model-based roughness measurements]{%
		\includegraphics[width=0.49\columnwidth]{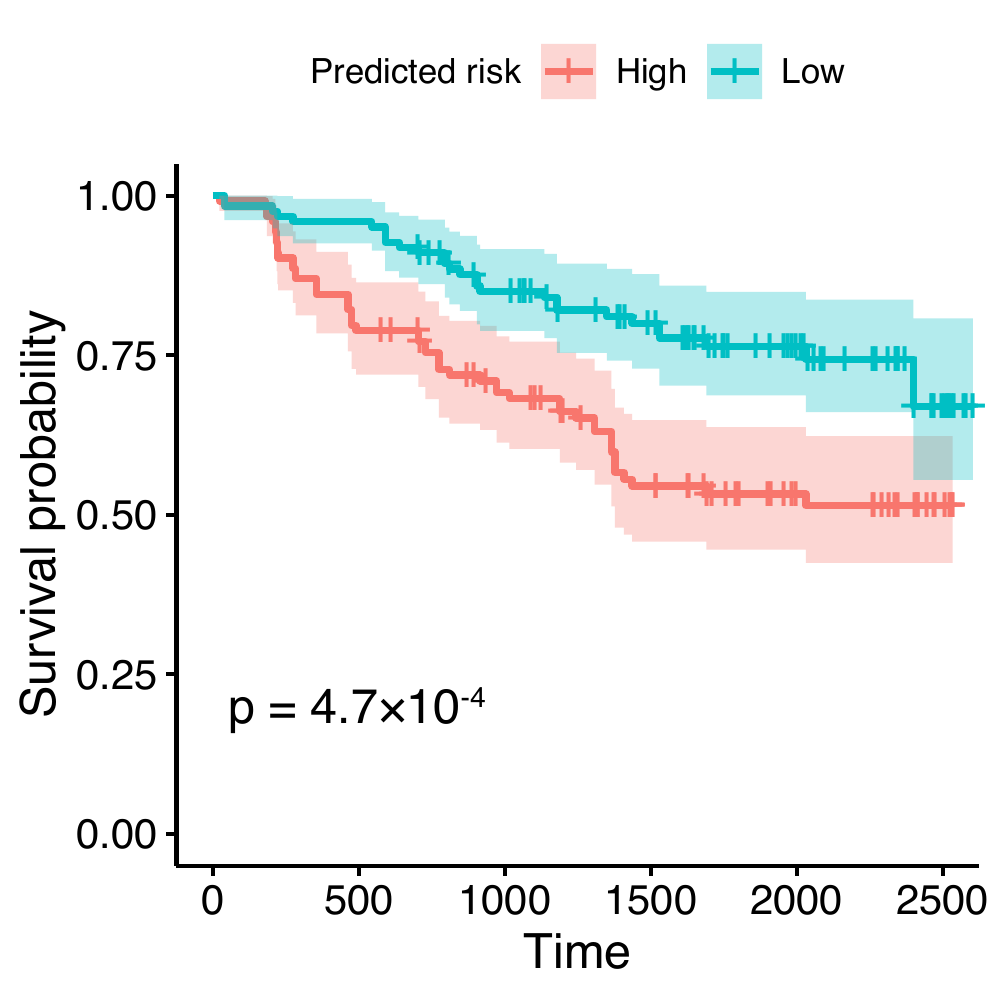}
	}
	\caption{Lung cancer case study: The Kaplan–Meier plots for the low and high-risk groups predicted by the leave-one-out cross-validation (LOOCV) via a multivariate Cox proportional hazards (CoxPH) model with survival time and vital status as responses and summary statistics of (a) distance-based and (b) model-based piecewise roughness measurements as the major predictors.}
	\label{f:fig_rapred}
\end{figure}

Lastly, we used the leave-one-out cross-validation (LOOCV) to evaluate the predictive performance of the above two CoxPH models, of which summary are shown in Table \ref{t:ra_cox} and \ref{t:neg2pos_cox}, respectively. For multiple pathology images that belong to each patient, we first trained a CoxPH model using all images from the rest patients. Next, we obtained a survival risk score for each test image. The survival risk score for the left-out patient was then calculated as the average survival risk score over all the associated images. After repeating this step for each patient of the $143$ NSCLC patient, we divided the patients into two equally sized groups (i.e. low and high-risk), choosing the median of patient-specific risk scores as the cutoff. Their corresponding Kaplan–Meier survival curves of those two groups are displayed in Figure \ref{f:fig_rapred}(a), where the predictors were the summary statistics of distance-based roughness measurements (i.e. Ra's). The log-rank test showed a significant difference between the two curves ($p$-value $=3.0\times10^{-6}$). The same LOOCV procedure was applied to evaluate the predictive performance of the CoxPH model that used model-based roughness features (i.e. transition probabilities). The Kaplan–Meier plot is shown in Figure~\ref{f:fig_rapred}(b). Again, the log-rank test showed a significant difference ($p$-value $= 4.7\times10^{-4}$) between the predicted high and low-risk groups.


\section{Conclusion} \label{conclusion}
A large amount of complex and comprehensive information about tumor aggressiveness and malignancy is harbored in the tumor shapes captured by pathology imaging. Recent advances in deep learning methods have provided plausible approaches for automatic tumor segmentation in medical images at large scale. Shape features proven with success in radiomics analysis, however, is no longer satisfying in pathology image. We proposed a framework to analyze the tumor shape in pathology images in this work, namely Bayesian LAndmark-based Shape Analysis (BayesLASA). The first contribution is that we developed the automatic landmark detection model for a polygonal chain under the Bayesian paradigm with improved accuracy and efficiency. It could also be extended to applications in various scenarios, where a sequence of discretization points could present the shape of an object. The second part of this work has proposed two types of new landmark-based features to characterize heterogeneous tumor boundary roughness, including distance-based and model-based approaches. Our study demonstrated the prognostic value of those features in an association study with lung cancer pathology images.

In the future, several extensions of our model are worth investigating. First, we could generalize the kernel used to measure the deviation between a polygonal chain and its landmark chain. For instance, using a squared exponential, Mat\'ern, or rational quadratic kernel will help us incorporate spatial dependence or desired smoothness. Thus, landmark identification and smoothness quantification could be jointly inferred. Moreover, we would like to extend our framework to the high-resolution pathology images of other cancer types, such as glioblastoma. It would be a promising research direction.






\newpage
\bibliographystyle{imsart-nameyear}
\bibliography{ref.bib}

\end{document}